\documentclass[reprint,eqsecnum,floats,aps,amsmath,amssymb,nofootinbib,prd,onecolumn, showpacs]{revtex4-1}

\usepackage{graphicx,physics}
\usepackage{amsmath,amssymb,mathtools,mathrsfs}
\usepackage{hyperref}
\usepackage{graphicx}
\usepackage{subfigure}
\usepackage{arydshln}
\usepackage{xcolor}
\usepackage{braket}
\usepackage{tensor}
\usepackage{enumitem,array,textcomp}

\begin{document}
	
	\title{Mart\'in-Benito--Mena Marug\'an--Olmedo prescription for the Dapor-Liegener model of loop quantum cosmology}
	\author{Alejandro Garc\'{\i}a-Quismondo}
	\email{alejandro.garcia@iem.cfmac.csic.es}
	\affiliation{Instituto de Estructura de la Materia, IEM-CSIC, Serrano 121, 28006 Madrid, Spain}
	\author{Guillermo  A. Mena Marug\'an}
	\email{mena@iem.cfmac.csic.es}
	\affiliation{Instituto de Estructura de la Materia, IEM-CSIC, Serrano 121, 28006 Madrid, Spain}
	
\begin{abstract}
Recently, an alternative Hamiltonian constraint for loop quantum cosmology has been put forward by Dapor and Liegener, inspired by previous work on regularization due to Thiemann. Here, we quantize this Hamiltonian following a prescription for cosmology proposed by Mart\'{\i}n-Benito, Mena Marug\'an, and Olmedo. To this effect, we first regularize the Euclidean and Lorentzian parts of the Hamiltonian constraint separately in the case of a Bianchi I cosmology. This allows us to identify a natural symmetrization of the Hamiltonian which is apparent in anisotropic scenarios. Preserving this symmetrization in isotropic regimes, we then determine the Hamiltonian constraint corresponding to a Friedmann-Lema\^itre-Robertson-Walker cosmology, which we proceed to quantize. We compute the action of this Hamiltonian operator in the volume eigenbasis and show that it takes the form of a fourth-order difference equation, unlike in standard loop quantum cosmology, where it is known to be of second order. We investigate the superselection sectors of our constraint operator, proving that they are semilattices supported only on either the positive or the negative semiaxis, depending on the triad orientation. Remarkably, the decoupling between semiaxes allows us to write a closed expression for the generalized eigenfunctions of the geometric part of the constraint. This expression is totally determined by the values at the two points of the semilattice that are closest to the origin, namely the two contributions with smallest eigenvolume. This is in clear contrast with the situation found for the standard Hamiltonian of loop quantum cosmology, where only the smallest value is free. This result indicates that the degeneracy of the new geometric Hamiltonian operator is equal to two, doubling the possible number of solutions with respect to the conventional quantization considered until now.
\end{abstract}


\maketitle

%%%%%

\section{Introduction}

Loop quantum gravity (LQG) is one of the most promising approaches to the construction of a quantum theory of gravity \cite{ALQG,Thiem}. It is a nonperturbative quantization of general relativity (GR) in 3+1 dimensions. In its canonical formulation, the basic variables are holonomies of the Ashtekar-Barbero connection along loops and fluxes of densitized triads across surfaces \cite{ALQG,AV2,BV1}. A key feature of the LQG program is that it employs a quantum representation which is not unitarily equivalent to the Fock representation of ordinary quantum field theory, but it is instead compatible with the background independence characteristic of GR. Besides, the Gauss, spatial diffeomorphism, and Hamiltonian constraints (which encode the information about the symmetries of GR in a triadic description) play a central role in this formulation. Following Dirac \cite{Dirac}, they are implemented as operators that annihilate the physical states.

The theoretical foundations of LQG have been explored during the three past decades and, although the full theory is not yet complete, some symmetry reduced sectors have already been studied in detail, as well as perturbations around them, pushing LQG closer to observationally falsifiable predictions \cite{FMO1,FMO2,CMMO,AAN1,AAN2,AAN3}. In particular, techniques inspired in LQG have proven successful in their application to cosmological systems in a complete and consistent manner. The field of research born from this merger of LQG and cosmology receives the name of loop quantum cosmology (LQC) \cite{BojLQC,AS,lqc3}. This field has undergone a major development lately, in great part owing to the large number of analyses carried out in homogeneous and isotropic models such as Friedmann-Lema\^itre-Robertson-Walker (FLRW) spacetimes (see e.g. Refs.  \cite{ABL,APS1,APS2,lqc3,MMO,APSV,SKL,PA}) and anisotropic Bianchi cosmologies (see e.g. Refs.  \cite{AW-E1,AW-E2,W-E,chiou1,chiou2,Bianchii,Bianchiii}). Some inhomogeneous cosmologies have also been studied, like for instance the Gowdy spacetimes \cite{GMM1,GM,GMM,MMW-E}. One of the most prominent results of LQC is the resolution of the initial big bang singularity, which is replaced by a quantum bounce in the Planckian regime once the quantum nature of gravity is taken into account. Before and after this bounce, the spacetime curvature decreases rapidly and GR soon becomes an extremely good approximation. Thus, this quantum bounce is often said to join two classical universes: a contracting prebounce branch and an expanding postbounce branch. Furthermore, this bouncing process is said to be deterministic in the sense that coherent states at large volumes remain peaked during their entire quantum trajectory \cite{Tav}, so that the peak follows an effective dynamics that is a very good approximation to the underlying quantum evolution \cite{APS2}. This singularity resolution (which is ultimately due to the discrete quantum geometry arising in the Planckian era) has been found to be general in isotropic and anisotropic cosmologies \cite{AS,Si,SS1,SS2,SS3} and has even been confirmed in inhomogeneous models \cite{FMT}. Indeed, several results point toward a general resolution of strong singularities in LQC, at least in very symmetric situations \cite{singh}.

One of the open issues in LQG is the choice of a quantum operator representing the Hamiltonian constraint and its regularization. Thiemann introduced a series of procedures to regularize this Hamiltonian in LQG, procedures which were later inherited by LQC. Nevertheless, the construction is affected by certain ambiguities \cite{TT,ALM,AAL}. Originally, one used physical arguments to try to overcome them. Lately, however, there has been a growing interest  in examining alternatives that might also result in viable physical pictures, comparing the resulting predictions, especially the possible modifications of the Planckian dynamics near the bounce. In particular, it would be enlightening to seek alternatives that can be adopted in full LQG without substantial modifications, investigate whether the resolution of cosmological singularities still holds for them, and discuss to what extent this resolution is maintained in LQG without symmetry reductions. 

In this respect, let us comment that there are fundamental differences between the regularization of the Hamiltonian constraint in full LQG and the regularization procedure that has been commonly used in standard LQC. Actually, the gravitational contribution to the Hamiltonian constraint in GR is composed by two pieces: the Euclidean and the Lorentzian parts. In full LQG each of them is quantized in a different manner. However, in the absence of spatial curvature, both terms are proportional. For this reason, in standard (flat) LQC, the Hamiltonian constraint has usually been written as proportional to its Euclidean part. Since the relation between LQG and LQC is not completely established yet (see e.g. Refs. \cite{Engle,BK}), it is not clear whether the cosmological dynamics arising in LQC reproduces the results of LQG, even at leading order. For this reason, it is convenient to study alternatives in cosmology which are closer to full LQG and, in this way, shed light on the question of the robustness of the bounce in the full theory.

One of the first attempts to address this issue was carried out by Yang, Ding, and Ma in Ref. \cite{YDM}. They considered an alternative Hamiltonian for LQC which is closer to the actual construction in LQG. The resulting Hamiltonian constraint turned out to be a fourth-order difference equation, instead of the second-order one of standard LQC. This departure from the conventional formalism of LQC has its origin in the separate quantization of the Euclidean and Lorentzian parts of the Hamiltonian that is carried out in LQG. Furthermore, a bounce mechanism was found with standard qualitative features, in the sense that there exist prebounce and postbounce branches that behave as solutions of classical GR at large volumes and are joined by a symmetric quantum bounce.

Much more recently, Dapor and Liegener \cite{DL1} used a nongraph-changing regularization \cite{GT} to obtain the Hamiltonian constraint within full LQG. In order to derive an effective description from it, they considered a set of complexifier coherent states, peaked on both holonomies and fluxes, and computed the expected value of the constraint, arriving at an effective Hamiltonian which coincides with that found in Ref. \cite{YDM}, in the isotropic and homogeneous limit and at leading order. In addition, in a later work \cite{DL2}, the strict regularization procedure that was originally devised for the Hamiltonian of LQG was directly applied in the context of LQC, obtaining a modified effective constraint which coincides with that proposed in Refs. \cite{YDM, DL1}\footnote{This alternative choice of constraint leads to a formalism of LQC that is sometimes called mLQC-I in the literature.}. By analyzing the resulting quantum dynamics, it was proven that the initial cosmological singularity was still resolved with this constraint, but there was a qualitative modification of the standard bounce. Indeed, it was shown that, even though this bounce still joins deterministically two large classical universes, there appears a large de Sitter era with an emergent cosmological constant (a similar phenomenon with the emergence of a cosmological constant had already been claimed to happen in the interior of black holes \cite{BH}). When evolution is parametrized by cosmic time, two distinct cases were encountered. Remarkably, both of them feature an asymmetric bounce, unlike in the study of Ref. \cite{YDM}. One of them joins a de Sitter contracting solution and a classical expanding universe, whereas the other joins a classical contracting universe with a de Sitter expanding phase. 

On the other hand, Li, Singh, and Wang \cite{Paramc1} developed a systematic study of the effective cosmological dynamics resulting from this very same Hamiltonian --to which we will refer as the \emph{modified Hamiltonian} throughout the rest of this paper. Using the Friedmann-Raychaudhuri equations and Hamilton's equations, and employing numerical and analytical tools, that work studied the case of massless and massive scalar fields. It was pointed out that the Friedmann-Raychaudhuri and the Hamiltonian descriptions are not related by a one-to-one correspondence, but rather two sets of Friedmann-Raychaudhuri equations are needed to cover the complete evolution of a universe. If one naively considers only one of the sets of equations, a symmetric bounce is obtained. However, it was shown that a symmetric bounce is not physically consistent, and so the bounce must be asymmetric, as it had already been suggested in Refs. \cite{DL1,DL2}. In principle, this asymmetry might result in significant phenomenological differences. In a subsequent work \cite{Paramc2}, the same authors extended their analysis to another possible choice of Hamiltonian constraint different from the standard one, obtaining a formalism that they named mLQC-II. Additionally, they studied and compared the effective dynamics of these alternatives to standard LQC for various inflationary potentials, concluding that they share a universal behavior, with a nonsingular inflationary epoch in which a phase of superinflation is followed by conventional inflation.

There have also been efforts to extract the consequences of this quantization ambiguity for observable quantities in cosmology. In this context, Agullo computed the power spectrum of the scalar perturbations using the modified Hamiltonian and compared his predictions with the results that had already been obtained in standard LQC \cite{Agullo}.

The aim of this paper is to construct a new formalism of LQC based in the modified Hamiltonian constraint by adopting for its quantum implementation, instead of the prescription proposed in Ref. \cite{APS2}, the one put forward by Mart\'in-Benito, Mena Marug\'an, and Olmedo in Ref. \cite{MMO}, named the MMO prescription after the initials of these three authors. The MMO prescription incorporates in homogeneous and isotropic LQC a natural symmetrization of the Hamiltonian that is inspired by the consideration of Bianchi I cosmologies, and which involves a special treatment of the signs of the components of the triad. This symmetrization leads to a quantum theory with some very appealing features. In the first place, the quantum analogue of the classical singularity is decoupled at the kinematical level, without the need of imposing additional conditions. Furthermore, the resulting theory has superselection sectors which are significantly simpler than those of other prescriptions. Thanks to this fact, it is then possible to get a closed expression for the generalized eigenfunctions of the gravitational Hamiltonian operator in standard LQC. Indeed, this expression turns out to be remarkably efficient from a computational viewpoint, allowing one to construct the eigenfunctions much faster and in a more precise manner. All of these features can be considered strengths of the MMO prescription. For this reason, it is interesting to verify whether such nice features persist when the MMO prescription is adopted for the quantum implementation of the modified Hamiltonian, rather than the standard one.

The rest of the paper is organized as follows. In Sec. \ref{sec:models}, we succinctly review the two cosmological models considered in the main body of this article: Bianchi I and flat FLRW spacetimes. We present the basics of their classical formulation and quantum kinematics, and illustrate the regularization of the Hamiltonian in these models by applying Thiemann's identities to the Euclidean part of the gravitational contribution. We then regularize the Lorentzian part in Sec. \ref{sec:fullH} in the case of Bianchi I cosmologies, expressing it in terms of holonomies and fluxes. Based on this computation, we identify the sign structure of the different terms of the modified Hamiltonian, introducing accordingly a natural symmetrization. We then consider the limiting situation of an isotropic regime to obtain the  regularized Lorentzian part of the Hamiltonian constraint in an FLRW spacetime. In Sec. \ref{sec:quantum}, we discuss the detailed implementation of the modified Hamiltonian as an operator (acting on the kinematical Hilbert space of the system) following the MMO prescription. Then, we compute the action of this operator on the basis of volume eigenstates, and determine the superselection sectors preserved  by this action. Finally, we consider generalized eigenstates of the modified Hamiltonian and discuss their degeneracy, obtaining an explicit closed expression that allows one to construct any of these eigenfunctions in terms of their data at two eigenvalues of the volume. We conclude in Sec. \ref{sec:conclusions} by summarizing and commenting our results.

%%%%%

\section{The Models}\label{sec:models}

In this section, we briefly review some relevant aspects of the two cosmological models that will be considered in this article: the Bianchi I spacetimes and the flat FLRW cosmologies. Given their simplicity and their applications to cosmology, these models have been extensively studied in LQC. For further details about these symmetry reduced models, both in GR and in LQC, one can consult e.g. Refs. \cite{kramer,APS2,MMO,AW-E1,Bianchii}. 

%%%%%

\subsection{Bianchi I cosmologies}\label{sec:BI}

Since we are interested in identifying a symmetrization of the modified Hamiltonian constraint that is valid not only in flat, homogeneous and isotropic cosmologies, but also in anisotropic scenarios, it is natural to consider a Bianchi I cosmological model as the starting point of our discussion. We begin by summarizing the classical aspects of the model that are relevant for our analysis.

Let us first describe the system in canonical Ashtekar-Barbero variables \cite{chiou1,AW-E1,Bianchii}. Although this definition usually involves a choice of a finite cell and a fiducial triad, it was shown in Ref. \cite{chiou2} that, incorporating a diagonal gauge, physical quantities do not depend on these choices provided that the canonical variables are defined appropriately. Then, for the sake of simplicity, we consider a diagonal Euclidean triad and particularize the discussion to spatial sections with a compact three-torus topology, so that it is natural to take as fiducial cell the whole of the $T^3$ section, with sides of coordinate length $2\pi$. In these circumstances, one arrives at the following Ashtekar-Barbero variables:
\begin{equation}
A_{a}^{i}=\dfrac{c^{i}}{2\pi}\delta^i_a,\qquad E^a_i=\dfrac{p_i}{4\pi^2}\delta^a_i,
\end{equation}
where $A_{a}^{i}$ denotes the $\mathfrak{su}(2)$-connection, $E^a_i$ the densitized triad, and spatial and SU(2) indices have been designated with lowercase Latin letters from the beginning and the middle of the alphabet, respectively. We will not use Einstein's summation convention unless explicitly stated. The nontrivial canonical Poisson brackets on the phase space of the model can be written in terms of the variables $c^i$ and $p_i$ as $\{c^i,p_j\}=8\pi G\gamma \delta^i_j$, where $G$ is Newton's gravitational constant and $\gamma$ is the Immirzi parameter. Notice that, in our diagonal gauge, internal $SU(2)$ indices and spatial indices can be identified, and we will do so in the following for convenience. Adopting circular coordinates $\{x^{i}\}=\{\theta,\sigma,\delta\}$, $x^i\in S^{1}$, the spacetime metric takes the form
\begin{equation}\label{lineel}
ds^2=-N^2dt^2+\dfrac{V^2}{4\pi^2}\sum_{i=\theta,\sigma,\delta}\dfrac{(dx^{i})^2}{p_{i}^2},
\end{equation}
where $N$ is the lapse function and $V=\sqrt{|p_\theta p_\sigma p_\delta|}$ is the physical volume of the Universe.

In homogeneous LQC, the configuration space is described using holonomies along straight edges. In the case of Bianchi I, these edges are oriented in the fiducial directions (labeled by $i=\theta,\sigma,\delta$) and have respective lengths $2\pi\mu_i\in\mathbb{R}$. In this way, one obtains the basic holonomies $h_i^{\mu_i}(c^i)=\text{exp}\{\mu_i c^i\tau_i\}$, where $\tau_i=-i\sigma_i/2$ (with $\sigma_i$ being the Pauli matrices) are the generators of the defining representation of $SU(2)$ and, as such, verify $[\tau_i,\tau_j]=\sum_{k}\epsilon_{ijk}\tau_k$. The phase space is completed with the triad fluxes through the fiducial rectangles formed by the edges of the holonomies. The area of the rectangle formed by edges in the two different fiducial directions $j$ and $k$, both orthogonal to the direction $i$, is proportional to $p_i$. Indeed, for a rectangular surface of this type with fiducial area $S_i$, the corresponding triad flux is $E(S_i)=S_i p_i/(4\pi^2)$.

The configuration algebra is given by sums of products of the matrix elements of the irreducible representations of the holonomies. As it is well known, this algebra is simply the algebra of almost periodic functions of the connection variables $c^i$, which is generated by the holonomy matrix elements $\prod \mathcal{N}_{\mu_i}(c^i)=\prod e^{i\mu_i c^i/2}$, where the product is over the three fiducial directions $i$ and $\mu_i$ is any real number. Employing the Dirac notation, each of the exponentials in these products will be represented upon quantization by a ket state $\ket{\mu_i}$. Denoting the tensor product of the states for the three different spatial directions by $\ket{\mu_\theta,\mu_\sigma,\mu_\delta}=\ket{\mu_\theta}\otimes\ket{\mu_\sigma}\otimes\ket{\mu_\delta}$, one can define the analogue of the space of cylindrical functions in LQG in the form  $\text{Cyl}_{\text{S}}^{\text{BI}}=\text{span}\left\{\ket{\mu_\theta,\mu_\sigma,\mu_\delta}\right\}$ \cite{GMM}.

The kinematical Hilbert space ${}^{(\text{BI})}\mathcal{H}^{\text{Kin}}=\otimes_i \mathcal{H}^{\text{Kin}}_i$ will then be given by the completion of $\text{Cyl}_{\text{S}}^{\text{BI}}$ with respect to the discrete inner product $\langle\mu_i|\mu_i'\rangle=\delta_{\mu_i,\mu_i'}$ in each direction. The states $\ket{\mu_\theta,\mu_\sigma,\mu_\delta}$ introduced above provide an orthonormal basis of this kinematical Hilbert space. Furthermore, the action of the fundamental operators on them is straightforward: they are eigenstates of the operator $\hat{p}_i$ (associated with triad fluxes across rectangles orthogonal to the $i$th fiducial direction), and its label is shifted by the operator $\hat{\mathcal{N}}_{\mu_i}$:
\begin{equation}
\hat{p}_i\ket{\mu_i}=4\pi \gamma l_p^2\mu_i\ket{\mu_i},\quad
\hat{\mathcal{N}}_{\mu_i'}\ket{\mu_i}=\ket{\mu_i+\mu_i'},
\end{equation}
where $l_p=\sqrt{G}$ is the Planck length in natural units. We set $\hbar=c=1$.

It has been argued that there should be a minimum coordinate length for the edges of holonomies in LQC \cite{APS2}, because the area spectrum is discrete in LQG and the limit of zero area cannot be reached. The most common prescription to take into account this consideration is the so-called $\bar{\mu}$ scheme or improved dynamics prescription. According to it, the minimum coordinate length for each direction is determined by the requirement that the edges of the holonomies form squares of minimum nonzero area. If the smallest nonzero eigenvalue allowed for this area in LQG  is denoted by $\Delta$, one then obtains the relation $\bar{\mu}_j\bar{\mu}_k|p_i|=\Delta$, with $i\neq j \neq k$, which yields $\bar{\mu}_i=\sqrt{\Delta |p_i|}/\sqrt{|p_jp_k|}$ \cite{AW-E1}. Since this value depends on the variables $p_i$ (for the three spatial directions), the shift produced by $\hat{\mathcal{N}}_{\bar{\mu}_i}$ becomes state dependent. To write the action of this holonomy operator in a simpler manner, we introduce an affine parameter $\lambda_i$ for each direction, defined by 
\begin{equation}
\lambda(p_i)= \textrm{sgn}(p_i) \frac{\sqrt{|p_i|}}{(4\pi \gamma l_p^2 \sqrt{\Delta})^{1/3}}.
\end{equation} 
Here, $\textrm{sgn}(\cdot)$ denotes the sign function. Then, we get $\partial_{\lambda_i}=2 \sqrt{|p_i|} \partial_{p_i}$, and we can represent $i\bar{\mu}_ic_i$ by the differential operator $8 \pi \gamma G \bar{\mu}_i \partial_{p_i}= |\lambda_j \lambda_k |^{-1}\partial_{\lambda_i}$ ($i\neq j \neq k$). Relabeling then the states of the $\ket{\mu_\theta,\mu_\sigma,\mu_\delta}$-basis, the fundamental holonomy operators have the action \cite{GMM}
\begin{equation}
\hat{\mathcal{N}}_{\pm\bar{\mu}_\delta}\ket{\lambda_\theta,\lambda_\sigma,\lambda_\delta}=\left| \lambda_\theta,\lambda_\sigma,\lambda_\delta\pm \frac{1}{2|\lambda_{\theta} \lambda_{\sigma}|} \right\rangle.
\end{equation}
We show only the case of one spatial direction, with similar actions for the others.

The next step is the representation of the Hamiltonian constraint, which is the only nontrivial constraint in a flat homogeneous cosmology for the introduced diagonal gauge. In full LQG, the gravitational part of the Hamiltonian constraint $H_{gr}$ consists of two parts: the Euclidean part $H_E$, which receives its name from the fact that it is the only one appearing in Euclidean gravity, and the Lorentzian part $H_L$. Explicitly, $H_{gr}(N)=N(H_E+H_L)$, with
\begin{eqnarray}
\label{HE}
H_E&=&\dfrac{1}{16\pi G}\int d^3x\, e^{-1}\epsilon\indices{^{ij}_k}E^a_iE^b_jF\indices{^k_{ab}},\\
\label{HL}
H_L&=&-\dfrac{1+\gamma^2}{8\pi G}\int d^3x\,e^{-1}E^a_iE^b_jK^{i}_{[a}K^{j}_{b]}.
\end{eqnarray}
Here, $e=\sqrt{|\text{det}(E)|}$, $K^i_a$ is the extrinsic curvature in triadic form, and $F\indices{^k_{ab}}$ is the curvature tensor of the Ashtekar-Barbero connection. Calling $\Gamma_a^i$ the spin connection compatible with the triad, we have that $\gamma K^{i}_a=A_a^i-\Gamma^{i}_a$ \cite{lqc3}. 

In flat cosmologies, the spin connection $\Gamma^{i}_a$ vanishes and, then, the Ashtekar-Barbero connection equals the triadic extrinsic curvature multiplied by the Immirzi parameter. Moreover, for homogeneous cosmologies, the extrinsic curvature is constant in each spatial section. Thus, in the case of the Bianchi I model, the curvature tensor becomes $F\indices{^{k}_{ab}}=\gamma^2\epsilon\indices{^{k}_{ij}}K^{i}_{[a}K_{b]}^{j}$. Hence, when contracted with $\epsilon\indices{^{ij}_k}$, it will provide a value for $H_L$ proportional to $H_E$. Indeed, for the case of flat homogeneous cosmologies, the gravitational Hamiltonian constraint can be solely written in terms of the Euclidean part as $H_{gr}(N)=-N H_E/\gamma^2$. For this reason, in LQC, it has been common to regularize the whole of the gravitational Hamiltonian as being proportional to the Euclidean part. In this way, one only needs to regularize the Euclidean Hamiltonian employing Thiemann's procedure. In the following, we will refer to this process as the ``standard regularization procedure.'' Let us now very briefly sketch this regularization.

Since $H_E$ depends on the curvature tensor, we begin by expressing it in terms of holonomies and fluxes via the following classical identity, valid in our diagonal gauge for Bianchi I:
\begin{equation}
{}^{(\text{BI})}F\indices{^i_{ab}}=-2\sum_{j,k}\text{tr}\left(\dfrac{h_{\square_{jk}}^{\bar{\mu}}-\delta_{jk}}{4\pi^2\bar{\mu}_j\bar{\mu}_k}\tau^i\right)\delta^j_a\delta^k_b,
\end{equation}
where the indices $i$, $j$ and $k$ are different. Besides, here $h_{\square_{jk}}^{\bar{\mu}}= h_j^{\bar{\mu}_j}h_k^{\bar{\mu}_k}(h_j^{\bar{\mu}_j})^{-1}(h_k^{\bar{\mu}_k})^{-1}$ and $(h_i^{\bar{\mu}_i})^{-1}$ is the inverse of the holonomy $h_i^{\bar{\mu}_i}$ along an edge of coordinate length $\bar{\mu}_i$ in the fiducial direction $i$. Then, computing $h_{\square_{jk}}^{\bar{\mu}}$ and introducing the result into the expression of the curvature tensor, one obtains that, for Bianchi I,
\begin{equation}
 {}^{(\text{BI})}F_{ab}^i=\dfrac{1}{\pi^2}\sum_{l}\epsilon_{abl}\dfrac{\sin \bar{x}_a}{\bar{\mu}_a}\dfrac{\sin \bar{x}_b}{\bar{\mu}_b}(\cos\bar{x}_a\cos\bar{x}_b\delta^i_l-\epsilon\indices{_{lb}^i}\cos\bar{x}_a\sin\bar{x}_b+\epsilon\indices{_{al}^i}\sin\bar{x}_a\cos\bar{x}_b),
\end{equation}
where $\bar{x}_i=\bar{\mu}_ic^i/2$, and spatial and $SU(2)$ indices are identified according to our conventions. Using this result, the Euclidean part of the gravitational Hamiltonian in a Bianchi I cosmology, that we will call $H_{E}^{\text{BI}}$, can be written as
\begin{equation}\label{HEBI}
H_{E}^{\text{BI}}=\dfrac{1}{16\pi G}\dfrac{1}{V}\sum_ip_i\dfrac{\sin 2\bar{x}_i}{\bar{\mu}_i}\sum_{j\neq i}p_j\dfrac{\sin 2\bar{x}_j}{\bar{\mu}_j},
\end{equation}
where we recall that $\bar{\mu}_i=\sqrt{\Delta |p_i|}/\sqrt{|p_jp_k|}$. In standard LQC, the gravitational Hamiltonian is given by the above expression up to a multiplicative factor. The representation of this Hamiltonian as an operator according to the MMO prescription will be briefly reviewed in the next section.

%%%%%

\subsection{FLRW cosmologies}\label{sec:flrw1}

Let us now discuss the isotropic case in which the three fiducial directions of our formalism for Bianchi I of the previous subsection display the same behavior and become equivalent, so that one attains a FLRW cosmology (for more details about the quantization of these cosmologies in LQC, see e.g. \cite{APS1,APS2,MMO}).

In this isotropic case, the Ashtekar-Barbero variables and their fundamental Poisson brackets reduce to
\begin{equation}
A_a^i=\dfrac{c}{V_0^{1/3}}\delta_a^i,\qquad E^a_i=\dfrac{p}{V_0^{2/3}}\delta_i^a,\qquad \{c,p\}=\dfrac{8\pi G\gamma}{3},
\end{equation}
where $V_0$ is the volume of the chosen finite fiducial cell (namely, $V_0=(2 \pi )^3$ for the considered toroidal sections). The above expressions are obtained with the fiducial Euclidean metric and the diagonal gauge, fixing in this way the spatial diffeomorphism and the gauge freedoms. The resulting LQC model is known to be independent of this choice of fiducial structures. We also notice the difference of a factor $1/3$ in the Poisson brackets of $c$ and $p$ with respect to the brackets of $c^i$ and $p_i$ in the Bianchi I model. This difference arises owing to the identification of the pairs  $(c^i,p_i)$ for the three spatial directions.

Following a procedure similar to that explained for Bianchi I, we next introduce holonomies along edges of coordinate length $V_0^{1/3}\mu$ and fluxes through squares formed by these edges. The configuration algebra is then the algebra of almost periodic functions of the connection variable $c$, generated by the holonomy matrix elements $\mathcal{N}_\mu(c)=e^{i\mu c/2}$. Upon quantization, these exponentials are represented by the states $\ket{\mu}$. The completion of $\text{Cyl}_{\text{S}}=\text{span}\{\ket{\mu}\}$ with respect to the discrete inner product yields the kinematical Hilbert space of the homogeneous and isotropic theory, $\mathcal{H}^{\text{Kin}}$. The action of the fundamental operators on the $\ket{\mu}$-basis is just $\hat{p}\ket{\mu}=(4\pi \gamma l_p^2\mu/3)\ket{\mu}$ and $\hat{\mathcal{N}}_{\mu'}\ket{\mu}=\ket{\mu+\mu'}$.

Adopting again the prescription of the improved dynamics in order to fix the coordinate length of the edges of the holonomies, so that they form rectangles with an area equal to the minimum nonzero eigenvalue allowed by LQG, we obtain that $\bar{\mu}=\sqrt{\Delta/|p|}$, which reproduces the result of Bianchi I particularized to the considered isotropic scenario. Introducing now an affine parameter $v$ so that the differential operator corresponding to $i\bar{\mu}c/2$ can be identified as $\partial_v$ \cite{APS2}, and relabeling the states $\ket{\mu}$ consequently, we get
\begin{equation}
\label{p}\hat{p}\ket{v}=(2\pi\gamma l_p^2\sqrt{\Delta})^{2/3}\text{sgn}(v)|v|^{2/3}\ket{v},\quad \hat{\mathcal{N}}_{\bar{\mu}}\ket{v}=\ket{v+1}.
\end{equation}
The parameter $v$ has a very clear geometrical interpretation: it is proportional to the physical volume. Indeed, particularizing the line element \eqref{lineel} to the analyzed isotropic case, the volume operator can be defined as $\hat{V}=\widehat{|p|}^{3/2}$, and hence $\hat{V}\ket{v}=2\pi\gamma l_p^2\sqrt{\Delta}\,|v|\ket{v}$.

Taking into account all these considerations, it is easy to realize that, in order to get the Euclidean part of the gravitational Hamiltonian for the isotropic FLRW model, it suffices to set $c_i= c$, $p_i= p$ and $\bar{\mu}_i=\bar{\mu}$ $\forall i$ in Eq. \eqref{HEBI}. We thus obtain
\begin{equation}
H_E=\dfrac{3}{8\pi G V}\left(\text{sgn}(p)|p|\dfrac{\sin(\bar{\mu}c)}{\bar{\mu}}\right)\left(\text{sgn}(p)|p|\dfrac{\sin(\bar{\mu}c)}{\bar{\mu}}\right).
\end{equation}
We keep this structure explicitly, in spite of the appearance of a squared sign, to facilitate the implementation of the MMO quantization prescription. 

%%%%%

\section{The full Hamiltonian constraint}\label{sec:fullH}

In this section, we will discuss the regularization of the Lorentzian part of the Hamiltonian constraint \eqref{HL}. This will allow us to construct the full gravitational Hamiltonian constraint by combining the result of this regularization with the regularized Euclidean part studied in Sec. \ref{sec:models}. Its particularization to the isotropic scenario will be quantized according to the MMO prescription in Sec. \ref{sec:quantum}.

%%%%%

\subsection{Lorentzian part in Bianchi I cosmologies}\label{sec:HLBI}

The regularization of the Lorentzian part \eqref{HL} employed to define the modified Hamiltonian constraint is based on the following classical identity, valid in full LQG:
\begin{equation}\label{K}
K_a^i=\dfrac{1}{8\pi G\gamma^3}\{A_a^i,\{H_E,V\}\}.
\end{equation}
Computing the Poisson brackets of the volume with the regularized Euclidean part for the Bianchi I model, one gets 
\begin{equation}\label{brack}
\{H_E^{\text{BI}},V\}=4\pi G\gamma\sum_{i}\text{sgn}(p_i)\sqrt{\bigg|\dfrac{p_jp_k}{p_i}\bigg|}\dfrac{\partial H_E^{\text{BI}}}{\partial c^i}=\dfrac{\gamma}{2}\sum_{i}p_i\dfrac{\sin \bar{x}_i}{\bar{\mu}_i}\sum_{j\neq i}\cos\bar{x}_j,
\end{equation}
where $i\neq j\neq k$. With these equations, it is straightforward to derive the regularized expression of the triadic extrinsic curvature for Bianchi I,
\begin{equation}\label{KBI}
{}^{(\text{BI})}K_a^i=\dfrac{1}{4\pi\gamma}\dfrac{\sin\bar{x}_i}{\bar{\mu}_i}\left(\sum_{j\neq i}\cos\bar{x}_j\right)\delta_a^i.
\end{equation}

The Lorentzian part of the gravitational Hamiltonian can then be written in Bianchi I as
\begin{equation}
H_L^{\text{BI}}=-\dfrac{1+\gamma^2}{2 G}\dfrac{\pi}{V}\sum_{i,j}p_ip_j \,{}^{(\text{BI})}K_{[i}^i {}^{(\text{BI})}K^j_{j]}.
\end{equation}
Inserting in this formula the regularized expression of ${}^{(\text{BI})}K_a^i$, one obtains
\begin{equation}\label{HLBI}
H_L^{\text{BI}}=-\dfrac{1}{64\pi G}\dfrac{1+\gamma^2}{\gamma^2}\dfrac{1}{V}\sum_ip_i\dfrac{\sin 2\bar{x}_i}{\bar{\mu}_i}\sum_{j\neq i}p_j\dfrac{\sin 2\bar{x}_j}{\bar{\mu}_j}\sum_{k\neq i}\cos 2\bar{x}_k\sum_{l\neq j}\cos 2\bar{x}_l.
\end{equation}
Finally, combining this result with the regularized Euclidean part of the Hamiltonian constraint \eqref{HEBI}, one arrives at the following formula for the gravitational Hamiltonian regularized according to the modified scheme:
\begin{equation}
H_{gr}^{\text{BI}}(N)=\dfrac{N}{16\pi G V}\sum_ip_i\dfrac{\sin\bar{\mu}_ic^i}{\bar{\mu}_i}\sum_{j\neq i}p_j\dfrac{\sin\bar{\mu}_jc^j}{\bar{\mu}_j}\left\{1-\dfrac{1+\gamma^2}{4\gamma^2}\sum_{k\neq i}\cos\bar{\mu}_kc^k\sum_{l\neq j}\cos\bar{\mu}_lc^l\right\}.
\end{equation}
It is important to emphasize that this modified Hamiltonian has a sign structure of the form 
\begin{equation}\label{signs}
\sum_{N=1,2} \left[ \sum_{i}\text{sgn}(p_i)F_i^{(N)}(|p|,c)\sum_{j\neq i}\text{sgn}(p_j) F_j^{(N)}(|p|,c) \right],
\end{equation}
where, for each value of the index $i$ and of the label $N=1,2$, the functions $F_i^{(N)}(|p|,c)$ depend only on the three connection variables $c^l$ and the norm of the triad variables $p_l$ (with $l=\theta, \sigma,\delta$), but not on the signs of $p_l$. Since the signs of the triad variables do not commute under the Poisson bracket with the connections in general, it seems natural to symmetrize the products of these variables in the quantization process. It seems natural as well to adopt the same symmetrization prescription when passing to the isotropic FLRW model, as a special case in which all spatial directions possess the same behavior (rather than rearranging the products to get factors that are quadratic in signs, and that would then equal the unit when evaluated in an isotropic situation). This choice of symmetrization and its preservation for FLRW cosmologies is the main core of the MMO prescription.  

%%%%%

\subsection{Lorentzian part in FLRW spacetimes: The isotropic case}

If we set $c_i= c$, $p_i= p$, and $\bar{\mu}_i= \bar{\mu}$ for all spatial directions $i$, as we did in Sec. \ref{sec:flrw1}, we obtain
\begin{equation}\label{Hgr}
H_{gr}(N)=\dfrac{3N}{8\pi GV}\left(\text{sgn}(p)|p|\dfrac{\sin\bar{\mu} c}{\bar{\mu}}\right)\left(\text{sgn}(p)|p|\dfrac{\sin\bar{\mu} c}{\bar{\mu}}\right)\left[1-\dfrac{1+\gamma^2}{4\gamma^2}\left(2\cos\bar{\mu}c\right)\left(2\cos\bar{\mu}c\right)\right],
\end{equation}
where we have maintained the sign structure in spite of the classical character of the expression, and we have obviated any reference to the fact that the modified Hamiltonian has been evaluated on FLRW cosmologies to simplify the notation. We can trivially cast this Hamiltonian as a difference of squares:
\begin{equation}\label{Hgr2}
H_{gr}=\dfrac{3N}{8\pi G V}\left\{\left(\text{sgn}(p)|p|\dfrac{\sin \bar{\mu}c}{\bar{\mu}}\right)\left(\text{sgn}(p)|p|\dfrac{\sin \bar{\mu}c}{\bar{\mu}}\right)-\dfrac{1+\gamma^2}{\gamma^2}\left(\text{sgn}(p)|p|\dfrac{\sin 2\bar{\mu}c}{2\bar{\mu}}\right)\left(\text{sgn}(p)|p|\dfrac{\sin 2\bar{\mu}c}{2\bar{\mu}}\right)\right\},
\end{equation}
where, we recall, $\bar{\mu}=\sqrt{\Delta/|p|}$. Notice that, for small $\bar{\mu}$, this Hamiltonian coincides with the one obtained for LQC via the standard regularization procedure. Indeed,
\begin{equation}
H_{gr}\approx-\dfrac{3N}{8\pi G\gamma^2V}\left(\text{sgn}(p)|p|\dfrac{\sin \bar{\mu}c}{\bar{\mu}}\right)\left(\text{sgn}(p)|p|\dfrac{\sin \bar{\mu}c}{\bar{\mu}}\right)
\end{equation}
when $\bar{\mu}c\ll 1$. Besides, as expected, in the (classical) limit $\bar{\mu}\to 0$ we recover the symmetry reduced gravitational Hamiltonian for flat homogeneous and isotropic cosmologies in GR,
\begin{equation}
    H_{gr}=-\dfrac{3N}{8\pi G\gamma^2}c^2\sqrt{|p|}.
\end{equation}
Hence, one expects to recover the features and results of standard cosmology and effective LQC in the limit of small minimum coordinate length.

Although the above remarks may appear to indicate that the inclusion of the Lorentzian part in the regularization procedure could be understood as a higher-order correction to the standard formalism, introducing no qualitative changes, this conclusion is not completely correct. In fact, we will show in Sec. \ref{sec:quantum} that the number of eigensolutions of this modified Hamiltonian seems to be doubled compared to those of the standard gravitational Hamiltonian of LQC.

%%%%%

\section{The quantum Hamiltonian}\label{sec:quantum}

In the quantum theory, the gravitational Hamiltonian constraint \eqref{Hgr} is represented by an operator acting on the Hilbert space of the system. It is well known that this representation is not unique and, in general, there exist nonequivalent ways of carrying out this quantization procedure which are \emph{a priori} valid. In this section, we will quantize the Hamiltonian constraint according to the MMO prescription. Once we have obtained the corresponding operator, we will compute its action on the volume eigenbasis and study its superselection sectors. We will conclude by discussing the generalized eigenstates of the gravitational Hamiltonian operator. 

%%%%%

\subsection{MMO quantization prescription}

In the MMO prescription, the factor ordering ambiguity that affects the quantum representation of the constraint is used to select a symmetric prescription that decouples the singular state $|v=0\rangle$ at the kinematical level, effectively removing this eigenvalue for all practical purposes from the discrete spectrum of the inverse volume operator. This symmetrization prescription was inspired by a thorough analysis of the Bianchi I cosmologies in LQC \cite{Bianchii,Bianchiii}, a fact that explains our discussion of those cosmologies in Secs. \ref{sec:BI} and \ref{sec:HLBI}. For such anisotropic spacetimes, the orientation of the triad (in the form of the sign of the triad variables) plays a fundamental role in LQC. Indeed, owing to the existence of three different spatial directions, the product of two of those signs is not necessarily equal to the identity, as it occurs in the case of FLRW cosmologies. Hence, this issue must be taken into account when symmetrizing the operator that represents the Hamiltonian constraint; otherwise, Bianchi I cosmologies could not be regarded as an immediate anisotropic generalization of the flat FLRW model. In brief, this prescription is based mainly on two rules for factor ordering ambiguities:
\begin{enumerate}[label=\roman*)]
\item Upon quantization, the products of powers of $\widehat{|p|}$ and $\widehat{1/\sqrt{|p|}}$ with holonomies and orientation signs are ordered adopting an algebraic symmetrization in the $|p|$-operators.
\item The terms involving the sign of $p$ are symmetrized as follows:
\begin{equation}
\sin n\bar{\mu} c\ \text{sgn}(p)\longrightarrow \dfrac{1}{2}\left\{\widehat{\sin n\bar{\mu}c}\ \widehat{\text{sgn}(p)}+ \widehat{\text{sgn}(p)}\ \widehat{\sin n\bar{\mu}c}\right\},
\end{equation}
for any integer $n$.
\end{enumerate}

Let us introduce the notation
\begin{equation}
\hat{\Omega}_{n\bar{\mu}}=\dfrac{1}{4i\sqrt{\Delta}}\left[\widehat{\dfrac{1}{\sqrt{|p|}}}\right]^{-1/2}\widehat{\sqrt{|p|}}\left\{(\hat{\mathcal{N}}_{n\bar{\mu}}-\hat{\mathcal{N}}_{-n\bar{\mu}})\ \widehat{\text{sgn}(p)}+\widehat{\text{sgn}(p)}\ (\hat{\mathcal{N}}_{n\bar{\mu}}-\hat{\mathcal{N}}_{-n\bar{\mu}}) \right\}\widehat{\sqrt{|p|}} \left[\widehat{\dfrac{1}{\sqrt{|p|}}}\right]^{-1/2},
\end{equation}
for any integer label $n$. Then, according to the above symmetrization rules and recalling that $\bar{\mu}=\sqrt{\Delta/|p|}$, it is straightforward to see that in the MMO prescription
\begin{equation}
|p|\,\text{sgn}(p)\ \dfrac{\sin\bar{\mu}c}{\bar{\mu}}\longrightarrow
\hat{\Omega}_{2\bar{\mu}}.
\end{equation}
Thus, we can write the Euclidean part of the gravitational contribution to the Hamiltonian constraint as
\begin{equation}
\hat{H}_E=\dfrac{3}{8\pi G}\left[\widehat{\dfrac{1}{V}}\right]^{1/2}\hat{\Omega}_{2\bar{\mu}}^2\left[\widehat{\dfrac{1}{V}}\right]^{1/2}.
\end{equation}
In these expressions,
\begin{equation}
\widehat{\dfrac{1}{V}}=\left[\widehat{\dfrac{1}{\sqrt{|p|}}}\right]^3
\end{equation}
and
\begin{equation}
\widehat{\dfrac{1}{\sqrt{|p|}}}=\dfrac{3}{4\pi G\gamma\sqrt{\Delta}}\widehat{\text{sgn}(p)}\widehat{\sqrt{|p|}}\left(\hat{\mathcal{N}}_{-\bar{\mu}}\widehat{\sqrt{|p|}}\hat{\mathcal{N}}_{\bar{\mu}}-\hat{\mathcal{N}}_{\bar{\mu}}\widehat{\sqrt{|p|}}\hat{\mathcal{N}}_{-\bar{\mu}}\right),
\end{equation}
which is known to be self-adjoint, and where $\widehat{\sqrt{|p|}}$ is defined from \eqref{p} in the sense of the spectral theorem \cite{Reed,Galindo}.

As can be seen by direct comparison in Eq. \eqref{Hgr2}, the Lorentzian part has in fact a form similar to that of the Euclidean part. Indeed, up to constant multiplicative factors, they are identical when $\bar{\mu}$ is substituted by $2\bar{\mu}$. Therefore, we are going to represent both parts of the Hamiltonian constraint by symmetric operators with the same structure. 
In this way, the operator representation of the Lorentzian part can be written as
\begin{equation}
\hat{H}_L=-\dfrac{3}{8\pi G}\dfrac{1+\gamma^2}{4\gamma^2}\left[\widehat{\dfrac{1}{V}}\right]^{1/2}\hat{\Omega}_{4\bar{\mu}}^2\left[\widehat{\dfrac{1}{V}}\right]^{1/2}.
\end{equation}

In conclusion, the full gravitational term in the Hamiltonian constraint operator for the MMO prescription is given by
\begin{equation}\label{HgrOp}
\hat{H}_{gr}(N)=\dfrac{3N}{8\pi G}\left[\widehat{\dfrac{1}{V}}\right]^{1/2}\left\{\hat{\Omega}_{2\bar{\mu}}^2-\dfrac{1+\gamma^2}{4\gamma^2}\hat{\Omega}^2_{4\bar{\mu}}\right\}\left[\widehat{\dfrac{1}{V}}\right]^{1/2}.
\end{equation}

%%%%%

\subsection{Action on the volume eigenbasis}

Since, with our symmetrization prescription, there are powers of the inverse volume operator both to the left and to the right of the gravitational Hamiltonian constraint, this operator is the first and the last to act on any state. Given that the inverse volume operator is, in turn, defined in terms of $\widehat{1/\sqrt{|p|}}$, it is ultimately a power of this operator that acts in the first and last places. From the definitions of $\widehat{\sqrt{|p|}}$ and $\hat{\mathcal{N}}_{\bar{\mu}}$, we get 
\begin{equation}
\widehat{\dfrac{1}{\sqrt{|p|}}}\ket{v}=\dfrac{3}{2(2\pi G \gamma \sqrt{\Delta})^{1/3}}|v|^{1/3}\bigg||v+1|^{1/3}-|v-1|^{1/3}\bigg|\ket{v}.
\end{equation}
In particular, it follows that the inverse volume operator annihilates the state of vanishing volume $\ket{v=0}$. Since, in addition, the action of this operator leaves invariant the orthogonal complement of $\ket{v=0}$, which we will call $\tilde{\mathcal{H}}^{\text{Kin}}$, the fact that the Hamiltonian constraint contains powers of the inverse volume at its left and right ends implies that the restriction of the quantum constraint to $\tilde{\mathcal{H}}^{\text{Kin}}$ is well defined. As a result, the singular state is decoupled and we can study the nontrivial solutions of the constraint in its orthogonal complement\footnote{We notice that $\tilde{\mathcal{H}}^{\text{Kin}}$ is simply the Cauchy completion of $\widetilde{\text{Cyl}}_{\text{S}}=\text{span}\{\ket{v},\,v\neq 0\}$ with respect to the discrete norm.}. In this sense, the quantum analogue of the classical singularity (i.e. the state $\ket{v=0}$) has been removed from the kinematical Hilbert space and, thus, the singularity is cured already at the kinematical level\footnote{\label{footmatter} We are assuming that the inclusion of a matter contribution to the constraint does not alter this conclusion.}.

With the kernel of the inverse volume operator already removed, we can proceed to the introduction of the densitized full Hamiltonian constraint operator, $\hat{\mathscr{H}}_{gr}$. Owing to the fact that physical states (that is, those that are annihilated by the Hamiltonian constraint operator) are, in general, not normalizable in the kinematical Hilbert space, it is natural to consider a larger space, namely, the algebraic dual of $\widetilde{\text{Cyl}}_{\text{S}}$ \cite{MMO}. In this dual space, it is possible to establish a one-to-one correspondence between any given element $\bra{\phi}$ annihilated by (the adjoint of) $\hat{H}_{gr}$ and the element $\bra{\phi'}=\bra{\phi}[\widehat{1/V}]^{1/2}$ annihilated by the (adjoint of) the densitized version of this constraint (see our comment in footnote \ref{footmatter})
\begin{equation}\label{Hgrdens}
\hat{\mathscr{H}}_{gr}(N)=\dfrac{3N}{8\pi G}\left\{\hat{\Omega}^2_{2\bar{\mu}}-\dfrac{1+\gamma^2}{4\gamma^2}\hat{\Omega}^2_{4\bar{\mu}}\right\}.
\end{equation}
In the following, we will refer to this densitized version when we mention the Hamiltonian constraint of the system, without further specifications.

According to the above expression, deducing the action of our Hamiltonian constraint operator amounts then to computing the action of $\hat{\Omega}^2_{n\bar{\mu}}$ for $n=2,4$. In order to do this, it is useful to define the functions
\begin{eqnarray}\label{g}
g(v)&=&\left\{\begin{array}{lc}
0& \text{if}\ v=0, \\
\bigg |\,\bigg |1+\dfrac{1}{v}\bigg |^{1/3}\!\!\!-\bigg |1-\dfrac{1}{v}\bigg |^{1/3}\bigg |^{-1/2}& \text{if}\ v\neq0,\end{array}\right.\\
\label{s}
s_{\pm}^{(n)}&=&\text{sgn}(v)+\text{sgn}(v\pm n),\\
\label{f}
f_{\pm}^{(n)}(v)&=&\dfrac{\pi G\gamma}{3}g(v)s_{\pm}^{(n)}(v)g(v\pm n).
\end{eqnarray}
In terms of these functions, the action of $\hat{\Omega}_{n\bar{\mu}}$ on the volume eigenbasis of $\tilde{\mathcal{H}}^{\text{Kin}}$ can be written in the compact form
\begin{equation}\label{omega}
\hat{\Omega}_{n\bar{\mu}}\ket{v}=-i\left[f_{+}^{(n)}(v)\ket{v+n}-f_{-}^{(n)}(v)\ket{v-n}\right].
\end{equation}
The action of $\hat{\Omega}_{n\bar{\mu}}^2$ is then
\begin{equation}
\hat{\Omega}_{n\bar{\mu}}^2\ket{v}=-f_{+}^{(n)}(v)f_{+}^{(n)}(v+n)\ket{v+2n}+\left\{\left[f_{+}^{(n)}(v)\right]^2+\left[f_{-}^{(n)}(v)\right]^2\right\}\ket{v}-f_{-}^{(n)}(v)f_{-}^{(n)}(v-n)\ket{v-2n}.
\end{equation}
In particular, we see that the action of the Euclidean part of the constraint operator may either preserve the label of the volume eigenstate or shift it in plus or minus four units, whereas the Lorentzian part produces shifts of plus or minus eight units, if any. Thus, given the discreteness of the volume representation, the action of the gravitational Hamiltonian constraint on any given state $\ket{v}$ results in an equation in finite differences that relates a total of five eigenstates: $\ket{v-8}$, $\ket{v-4}$, $\ket{v}$, $\ket{v+4}$, and $\ket{v+8}$. 

%%%%%

\subsection{Superselection sectors}

In standard isotropic LQC and with the prescriptions of Ref. \cite{APS2}, the gravitational Hamiltonian constraint leaves invariant the Hilbert spaces with support on discrete lattices of step four. In addition, these lattices turn out to be superselection sectors inasmuch as they are also preserved by the whole Hamiltonian constraint which includes the matter term (typically that corresponding to a scalar field; see in general footnote \ref{footmatter}), as well as by the relevant physical observables \cite{ABL,APS1,APS2,ACS,KL1,KL2}. On the other hand, the superselection sectors for the MMO prescription in standard LQC are even simpler. Owing to the special choice of factor ordering in the symmetrization of the constraint, the positive and negative semilattices decouple under the action of the Hamiltonian constraint \cite{MMO}. As a consequence, the superselection sectors are given by the Hilbert spaces with support on discrete semilattices of step four, instead of entire lattices. In this way, the MMO prescription results in simple superselection sectors, a fact which is regarded as an advantage over other quantization prescriptions. In this subsection, we will analyze whether this property of the MMO prescription still holds when the Lorentzian part of the Hamiltonian constraint is introduced in the regularization procedure.

Given that the operator representation of the Lorentzian part has the same structure as its Euclidean counterpart but produces shifts that are twice as large, the full Hamiltonian constraint leaves invariant the Hilbert spaces with support on any discrete lattice of step four. Then, at the very least, the resulting superselection sectors will not be more involved than those usually found in standard isotropic LQC. Actually, they turn out to be simpler again. This is due to the fact that the coefficients $f_{\pm}^{(n)}(v)f_{\pm}^{(n)}(v\pm n)$ have some special properties. Indeed, recalling their definition \eqref{f}, it is straightforward to verify that they satisfy
\begin{equation}\label{iden}
\begin{gathered}
f_{+}^{(n)}(v)f_{+}^{(n)}(v+ n)=0\quad \forall\, v\in[-2n,0),\\
f_{-}^{(n)}(v)f_{-}^{(n)}(v- n)=0\quad \forall\, v\in(0,2n].
\end{gathered}
\end{equation}
These identities show that the lattices split into two under the action of the Hamiltonian constraint \eqref{Hgrdens}, i.e., the positive and negative semiaxes decouple as well. Hence, the superselection sectors are Hilbert spaces with support on the positive or negative discrete semilattices. More precisely, the superselection sectors $\mathcal{H}_\varepsilon^{\pm}$ are the Cauchy completion with respect to the discrete norm of $\text{Cyl}_{\varepsilon}^{\pm}=\text{span}\{\ket{v}, \, v\in\mathcal{L}_\varepsilon^\pm\}$, where $\mathcal{L}_\varepsilon^\pm$ are the following semilattices of step four:
\begin{equation}
    \mathcal{L}_\varepsilon^\pm=\{v=\pm(\varepsilon+4n),\,\varepsilon\in(0,4],\, n\in\mathbb{N}\}.
\end{equation}

The support on semilattices, characteristic of these superselection sectors, is a consequence of the identities \eqref{iden}, which in turn follow from the special combinations of signs dictated by the symmetrization chosen for the Hamiltonian constraint operator. Therefore, in conclusion, the simple superselection sectors that we have found ultimately come from the special treatment of the orientation of the triad, which is a key feature of the MMO prescription. 

%%%%%

\subsection{Generalized eigenfunctions}

In standard LQC, the quantum Hamiltonian constraint has generally been found to admit self-adjoint extensions with a spectrum that has a nonempty continuous part. Assuming that these properties are also exhibited by the quantum Hamiltonian \eqref{Hgrdens}, assumption which is supported at least in part by previous investigations (see Ref. \cite{DL2}), we now focus our attention on this continuous spectrum and analyze the generalized eigenfunctions. The generalized eigenstates will be denoted
\begin{equation}\label{geneig}
\ket{e^\varepsilon_\lambda}=\sum e^\varepsilon_\lambda(v)\ket{v},\qquad \left[\hat{\Omega}^2_{2\bar{\mu}}-\dfrac{1+\gamma^2}{4\gamma^2}\hat{\Omega}^2_{4\bar{\mu}}\right]\ket{e^\varepsilon_\lambda}=\lambda\ket{e^\varepsilon_\lambda},
\end{equation} 
with eigenvalue of the Hamiltonian operator equal to $3N\lambda/8\pi G$. In this expression, the sums are taken over $v\in \mathcal{L}_\varepsilon^\pm$, with $\varepsilon$ fixed. Let us consider, for instance, the superselection sector $\mathcal{H}^+_\varepsilon$ corresponding to the semilattice $\mathcal{L}_\varepsilon^+$.

By projecting the second equality in Eq.\eqref{geneig} on $| v \rangle$, we typically find a relation between the values of the eigenfunction at five points of the semilattice. Remarkably, if we consider the closest point to the origin, namely $v=\varepsilon$, the relation involves only the values of the eigenfunction at three points: $\varepsilon$, $\varepsilon+4$, and $\varepsilon+8$. This is so because the coefficients corresponding to the two other possible points, that lie on the negative semilattice, turn out to vanish, as we discussed above. Hence, if we fix the eigenfunction at the two first points, $e^\varepsilon_\lambda(\varepsilon)$ and $e^\varepsilon_\lambda(\varepsilon+4)$, the third point becomes immediately fixed by the requirement that $\ket{e^\varepsilon_\lambda}$ be an eigenstate of the quantum Hamiltonian. If we move one step to the right (i.e. to the point that is the second-to-closest to the origin), only one of the five possible coefficients will vanish, and we will extract a linear relation between four values of the eigenfunction. Since three of them are already known, the fourth is fixed. This argument can be continued recursively \emph{ad infinitum}, and the conclusion is that the eigenfunction is determined at every point of the semilattice once we fix its two first values.

It is worth emphasizing that this result differs from the one obtained in standard LQC. With the usual regularization of the Hamiltonian adopted until now, there is only one piece of data left free, which can be identified with the value at the smallest eigenvalue of the volume in the superselection sector, and that can be absorbed up to a phase by a suitable (generalized) normalization of the eigenfunction \cite{MMO}. Indeed, in that regularization, the Hamiltonian is entirely determined by the Euclidean part. Hence, in principle, its action only relates three eigenvolumes, but at the end of the semilattice, $v=\varepsilon$, one of the contributions vanishes owing to the symmetrization of the sign factors, reducing the action to two points. If we provide the value at the smallest volume, then the value at the other point gets fixed, and from there on we can determine the rest of the values on the semilattice by means of the linear eigenvalue equation associated to the gravitational Hamiltonian operator. In conclusion, in the process of considering the Lorentzian part of the Hamiltonian constraint, we have increased the number of solutions. For this reason, this part cannot be regarded just as a mere higher-order corrective term that modifies the solutions that we had got previously. An entirely new class of eigensolutions arise from our modified regularization procedure in LQC. This is related with the appearance of the de Sitter solution of constant Planckian curvature in the Dapor-Liegener model, analyzed e.g. in Refs. \cite{DL2,Agullo,Paramc1,Paramc2,Haro}.

In order to complete this discussion, we will now derive a closed expression that allows one to compute any generalized eigenfunction (which we are here assuming that exists) at any point of the semilattice, in terms of the values at the two points of smallest volume. To write this expression in a more compact manner, it is useful to introduce the following definitions:
\begin{eqnarray}\label{defs1}
F^{0}_\lambda(v) &=&\dfrac{4\gamma^2}{1+\gamma^2}\dfrac{\lambda-\left\{\left[f_{+}^{(2)}(v)\right]^2+\left[f_{-}^{(2)}(v)\right]^2\right\}}{f_{-}^{(4)}(v+8)f_{-}^{(4)}(v+4)}+\dfrac{\left[f_{+}^{(4)}(v)\right]^2+\left[f_{-}^{(4)}(v)\right]^2}{f_{-}^{(4)}(v+8)f_{-}^{(4)}(v+4)},\\
F^{\pm 4}(v)&=& \dfrac{4\gamma^2}{1+\gamma^2}\dfrac{f_{\mp}^{(2)}(v\pm 4)f_{\mp}^{(2)}(v\pm 2)}{f_{-}^{(4)}(v+8)f_{-}^{(4)}(v+4)},\label{defs2}\\
F^{-8}(v)&=& -\dfrac{f_{+}^{(4)}(v-8)f_{+}^{(4)}(v-4)}{f_{-}^{(4)}(v+8)f_{-}^{(4)}(v+4)}.\label{defs3}
\end{eqnarray}
Actually, in the relation between the eigenfunction at five consecutive points centered around $v$, the above functions appear as the coefficients of the eigenfunction at $v$, $v\pm 4$, and $v-8$, respectively (after normalizing the coefficient of the eigenfunction at $v+8$ to one).

Employing the functions \eqref{defs1}-\eqref{defs3}, the value of a generalized eigenfunction at any finite point of the semilattice $\mathcal{L}_\varepsilon^+$ can be written in the form
\begin{equation}
e^\varepsilon_\lambda(\varepsilon\!+\!4n)=\!\!\!\sum_{m=0,1}\sum_{O(m\to n)}\left\{\prod_{\{r_m\}}\!F^{+4}[\varepsilon\!+\!4(r_m\!-\!1)]\!\prod_{\{s\}}\!F^0_\lambda[\varepsilon\!+\!4s]\!\prod_{\{t\}}\!F^{-4}[\varepsilon\!+\!4(t\!+\!1)]\!\prod_{\{u\}}\!F^{-8}[\varepsilon\!+\!4(u\!+\!2)]\right\}\,e^\varepsilon_\lambda (\varepsilon\!+\!4m).
\end{equation}
In this expression, $O(p\to q)$ denotes the set of paths containing jumps of one, two, three, or four units that connect two points $p$ and $q$ on a semilattice of step 1. These paths will be composed by intermediate points between $p$ and $q$. Then, we call $\{r\}$, $\{s\}$, $\{t\}$, and $\{u\}$ the subsets of these intermediate points that, in the considered path, are followed by jumps of one unit, two units, three units, and four units, respectively. 

It is important to notice the distinction between the set of intermediate points followed by a one-unit step in the case of paths starting at the integer 0 (i.e., $\{r_0\}$) from the same set in the case of paths starting  at the integer 1 (i.e., $\{r_1\}$). The difference between both sets lies in the fact that $0\not\in\{r_0\}$, whereas $1$ may belong to $\{r_1\}$. If $0$ did belong to $\{r_0\}$, there would be a term connecting $\varepsilon$ and $\varepsilon+4$ which does not appear in the direct computation. In the opposite case, one would conclude that $e^\varepsilon_\lambda(\varepsilon)$ and $e^\varepsilon_\lambda(\varepsilon+4)$ are not independent of each other, reducing the number of available free pieces of data back to one. Bearing this particularity in mind, the expression that we have deduced should allow us to compute the generalized eigenfunctions associated to a certain eigenvalue at any point of the considered semilattice by the determination of all the paths that connect the integers 0 and 1 with any other integer number.

%%%%%

\section{Conclusions and discussion}\label{sec:conclusions}

After almost two decades in which the field of LQC has developed intensively, there exists nowadays a growing interest in revisiting some aspects of its foundations and discuss thoroughly the issue of the possible mathematical ambiguities (affecting, for instance, the regularization procedure followed to define the quantum Hamiltonian constraint), ambiguities which may lead to new modified formalisms. Some of these alternatives may, in turn, allow the construction of viable physical descriptions which depart from the standard picture of the resolution of cosmological singularities encountered in the conventional formalism of LQC (e.g. by introducing modifications to the Planckian dynamics). For this reason, it is enlightening to analyze whether the qualitative big bounce scenario is robust when these ambiguities are taken into account. In particular, this is especially interesting in the case where the techniques employed in those alternative formalisms are closer to full LQG, since the relation between LQC and LQG is not yet fully understood and, hence, one may question the extent to which LQC is able to capture the full cosmological dynamics of LQG. Therefore, it is important to study alternative loop quantizations for cosmology that follow more faithfully the quantization program of LQG, and compare their physical predictions with those of standard LQC.

Recently, Dapor and Liegener proposed an alternative of this type \cite{DL1}. Indeed, they computed the gravitational part of the Hamiltonian constraint operator in full LQG and its expected value on complefixier coherent states that represent homogeneous and isotropic cosmologies. As a result, they obtained an effective Hamiltonian which coincides at leading order, in a semiclassical expansion, with a Hamiltonian that had already been considered in the literature by Yang, Ding, and Ma \cite{YDM}. It has also been shown \cite{DL2} that this precise Hamiltonian can be obtained by regularizing separately the Euclidean and Lorentzian parts of the gravitational Hamiltonian constraint. In the past months, there has been a considerable number of works devoted to the analysis of this modified Hamiltonian and the physics that results from  it. Recurring to an effective description of the associated cosmological dynamics, it has been shown that the initial singularity is replaced by a quantum bounce that joins deterministically a contracting prebounce branch and an expanding postbounce branch. However, there is a major departure from standard LQC: while one of the branches remains classical at large volumes, the other is a de Sitter universe with constant Planckian curvature.

Actually, the regularization procedure of the Hamiltonian constraint is not the only source of ambiguities. Indeed, different quantization prescriptions, motivated mainly by different choices in the factor ordering, have been proposed in the literature of standard LQC (see, for instance, Refs. \cite{APS2,MMO}). In this paper, we have investigated whether the nice properties of the MMO prescription in standard LQC still hold in the  modified formalism of Dapor and Liegener. In particular, we have addressed the following questions: i) Does the quantum analogue of the classical singularity still decouple at the kinematical level? ii) Does the prescription still result in superselection rules that pick out simpler superselection sectors? iii) Does this fact allow us to find closed expressions for the generalized eigenfunctions of the gravitational part of the Hamiltonian constraint? iv) If so, how many free pieces of data are available in the construction of these generalized eigenfunctions? Notice that, for each given eigenvalue, the number of such pieces should determine the degeneracy of the quantum gravitational Hamiltonian. 

With these objectives, we have started our analysis by considering a Bianchi I cosmology, in order to identify the natural symmetrization that corresponds to the MMO prescription and that becomes apparent in anisotropic spacetimes. In this setting, we have regularized the Euclidean part of the gravitational Hamiltonian in the usual manner, and the Lorentzian part in a very similar way, using the appropriate Thiemann's identities. Once the full gravitational Hamiltonian has been computed, we have identified the sign structure that appears thanks to the fact that the signs of different components of the triad need not be equal. Preserving this sign structure, we have passed to the isotropic case in order to analyze FLRW spacetimes. 

Next, we have implemented the MMO prescription to quantize the modified Hamiltonian of the Dapor-Liegener model for FLRW cosmology. We have shown that, owing to the factor ordering characteristic of the prescription, this modified Hamiltonian annihilates quantum mechanically the singular state of vanishing volume and leaves invariant its orthogonal complement. Consequently, since we are interested in nontrivial solutions to the Hamiltonian constraint, we have restricted our analysis to the orthogonal complement of the singular state. In this sense, the singularity has been removed, allowing us to densitize the constraint in the algebraic dual of the kinematical Hilbert space that has support on the aforementioned orthogonal complement. The action of this Hamiltonian on the basis of eigenstates of the volume operator turns out to provide a fourth-order quantum difference equation (that is, an equation that relates a total of five volume eigenstates), in contrast with the second-order difference equation that appears in standard LQC, or the differential equation from the Wheeler-de Witt theory.  In fact, the modified Hamiltonian, when acting on a state labelled by a certain eigenvalue of the volume operator $v$, produces shifts of four or eight units in its label (thus connecting $v$ with $v\pm 4$ or $v\pm 8$). As a result, this quantum Hamiltonian leaves invariant lattices of step four. Nonetheless, owing to the properties of the MMO prescription adopted in the construction of this operator, it leaves invariant smaller sets. Indeed, the positive and negative semilattices decouple, resulting in superselection sectors that are semilattices of step four.

The fact that there is a point in these semilattices with minimum (and nonvanishing in absolute value) volume eigenvalue allows us to write a relation between the \emph{three} first points by acting with the modified Hamiltonian constraint on the first one (the other two possible contributions, which would connect with the semilattice with opposite orientation, happen to vanish, explaining the decoupling). This has relevant consequences for the possible  generalized eigenfunctions of the modified Hamiltonian. Indeed, if we take fixed values for the eigenfunction at the two first points, the value at the third point of the semilattice is immediately determined by the eigenvalue equation. If we then displace the action of the constraint to the second point, \emph{four} points will be involved in the resulting relation. Given that the three first values are already fixed by now, this condition determines the fourth one. Note that we can extend this argument recursively to any finite point of the semilattice. In this way we conclude that there are only \emph{two} free pieces of data available to determine the whole generalized eigenfunction, which are given by its two first values. This conclusion differs from the situation found in the standard case. In this modified formalism, an extra free piece of data appears, indicating an increase in the degeneracy of the quantum Hamiltonian, owing to the introduction of its Lorentzian part and the subsequent regularization. 

Finally, we have also shown that it is not only possible to obtain the generalized eigenfunctions at each point recursively but, moreover, this can be done using a closed expression, which solely depends on the two first values of the eigenfunction in the considered superselection sector. Note that this is only feasible because the generalized eigenfunctions of the quantum Hamiltonian have support on a single semilattice, which contains no potential singularity. Furthermore, this behavior does not result from imposing a certain condition, such as a boundary condition. It is a characteristic feature of the functional properties of the gravitational Hamiltonian that follows directly from the specific treatment of the signs of the components of the triad adopted in the MMO prescription. 

In conclusion, the analysis presented in this work allows us to confirm that the MMO prescription does retain the appealing properties that characterize it in LQC, even after the introduction of a modified regularization procedure for the gravitational part of the Hamiltonian constraint.

\section*{Acknowledgements}
The authors are grateful to Beatriz Elizaga Navascu\'es for discussions. This work has been partially supported by Grant No. CSIC JAEINT18\texttt{\char`_}EX\texttt{\char`_}0226, and by Project. No. MINECO FIS2014-54800-C2-2-P and Project. No. MICINN FIS2017-86497-C2-2-P from Spain.

\end{document}